\documentclass[12pt,onecolumn,journal]{IEEEtran}

\newtheorem{theorem}{Theorem}
\newtheorem{prop}{Proposition}
\newtheorem{corol}{Corollary}
\newtheorem{defin}{Definition}

\newtheorem{remark}{Remark}
\newtheorem{example}{Example}

\newtheorem{assum}{\textbf{A}\hspace{-0.16cm}}

\newcommand{\Exp}{\ensuremath{\mathbb{E}}}

\usepackage{times}
\usepackage{amsmath}
\usepackage{amssymb}
\usepackage{epsfig}
\usepackage{float}
\usepackage{multirow}
\usepackage{graphicx}
\usepackage{cite}
\renewcommand{\QED}{\QEDopen}
\begin{document}
\centerfigcaptionstrue
\title{\Large{A Note on Comparison of Error Correction Codes}}
\author{\large ~Dejan~V.~Djonin\\
\thanks{The author is with Dyaptive Systems, Inc., Vancouver, BC and adjunct professor at the Department of Electrical and Computer Engineering,
University of British Columbia, 2356 Main Mall, Vancouver, BC, V6T
1Z4, Canada, e-mail:~ddjonin@ece.ubc.ca.}}
\maketitle
\thispagestyle{empty}
%
%
\begin{abstract}
Use of an error correction code in a given transmission channel can
be regarded as the statistical experiment. Therefore, powerful
results from the theory of comparison of experiments can be applied
to compare the performances of different error correction codes. We
present results on the comparison of block error correction codes
using the representation of error correction code as a linear
experiment. In this case the code comparison is based on the Loewner
matrix ordering of respective code matrices. Next, we demonstrate
the bit-error rate code performance comparison based on the
representation of the codes as dichotomies, in which case the
comparison is based on the matrix majorization ordering of their
respective equivalent code matrices.
\end{abstract}
%
%
%
%
%
%
\section{Introduction}\label{sec:introduction}
%
%
In this section we will review the basic concepts of statistical
experiments that will later be used to establish ordering relation
between error correction codes employed in a given communication
channel. Following\cite{Torgersen}, a \textit{statistical
experiment} is defined as a pair $\mathcal{E} = (\mathcal{X},
P_{\theta}; \theta \in \Theta)$ where $\mathcal{X}$ is a measurable
\textit{sample space}, $P_{\theta}$ is a probability measure on
$\mathcal{X}$ for each $\theta \in \Theta$ and $\Theta$ is a
\textit{parameter set}. Let the decision rule $y = \sigma(x)$ of
experiment $\mathcal{E}$ be defined as a mapping from an element of
the sample space $x \in \mathcal{X}$ to the element of the decision
space $y \in \mathcal{Y}$. Further, one can define the loss function
$\mathcal{L}(\theta,y)$ of choosing the decision $y \in \mathcal{Y}$
when the true state of the parameter is $\theta \in \Theta$. The
reader is referred to\cite{Torgersen} for a thorough treatment of
statistical experiments, loss functions, with numerous illustrative
examples. We will next introduce the notation used throughout the
paper, followed by the definition and properties of deficiency
distance between experiments.

\textit{Notation:} We will use $a'$ to denote the transpose of
vector $a$, $A^{+}$ denotes the Moore-Penrose pseudoinverse of the
matrix $A$, $\#S$ denotes the cardinality of set $S$. $A \wedge B$
and $A \vee B$ denote the element-wise minimum and maximum of two
equal dimensional matrices $A$ and $B$, respectively. Let $A(i,:)$
and $A(:,i)$  denote the $i$-th row and $i$-th column of the matrix
$A$, respectively. Range $\textit{range}(B)$ of the matrix $B$ is
the space spanned by the columns of the matrix $B$. Let $e_i$ be
$[0,0,\ldots,0,1,0,\ldots,0]$ where $1$ is at the $i$-th position of
vector $e_i$. Let $\textrm{conv}(a,b) = \{\alpha a + (1-\alpha)b |
\alpha \in [0,1] \}$ for equal dimensional vectors $a$ and $b$. Let
$\mathcal{M}_{nm}$ be the set of row-stochastic $n \times m$
dimensional matrices, i.e. these matrices have positive elements
whose rows sum to $1$. Let indicator function $I_{\{x\}}$ be equal
to 1 if $x$ is true and $0$ otherwise. $N(\mu, \Sigma)$ denotes the
multivariate normal random variable with mean $\mu$ and covariance
$\Sigma$. $\|f\|$ denotes the total variation of function $f(x)$
given by $\int |f(x)| dx$. $\Exp_{u}$ is the expectation with
respect to (w.r.t) random variable $u$.

Le Cam has introduced in~\cite{LeCam} the following definition of
the {\em deficiency distance} between two experiments:
\begin{defin}
\label{defin:deficiency} \textit{(Deficiency Distance)} Experiment
$\mathcal{E} = (\mathcal{X}, P_{\theta}; \theta \in \Theta)$ is
$\epsilon$-deficient w.r.t experiment $\mathcal{F} = (\mathcal{Y},
Q_{\theta}; \theta \in \Theta)$ if for each $\Theta_{0} \subset
\Theta$, and each decision rule $\rho$ of experiment $\mathcal{F}$,
there exists a decision rule $\sigma$ of experiment $\mathcal{E}$
such that
\begin{equation}
\int P_{\theta}(x) \mathcal{L}(\theta,\sigma(x)) dx \leq \int Q_{\theta}(y)
\mathcal{L}(\theta,\rho(y)) dy + \epsilon_{\theta} \max_{y}
\mathcal{L} (\theta,y)
\end{equation}
when $\theta \in \Theta_0$. In short form, the deficiency distance is denoted as $\epsilon
= \delta(\mathcal{E},\mathcal{F})$.
\end{defin}

The deficiency distance can plausibly be interpreted as follows: the
deficiency of experiment $\mathcal{E}$ w.r.t the experiment
$\mathcal{F}$ is the upper bound on the difference in the risk
functions\footnote{Expression $\int P_{\theta}(x)
\mathcal{L}(\theta,\sigma(x)) dx$ is commonly referred as the {\em
risk function} of the experiment $\mathcal{E} = (\mathcal{X},
P_{\theta}; \theta \in \Theta)$ for a given loss function $L$ and
decision rule $\sigma$.} between experiments $\mathcal{E}$ and
$\mathcal{F}$ for any choice of the a priori knowledge of the
unknown parameter $\theta$ and some decision rule on $\mathcal{E}$.
In general, therefore, the deficiency distance is \textit{not
symmetric}, i.e. $\delta(\mathcal{E},\mathcal{F}) \neq
\delta(\mathcal{F},\mathcal{E})$.

Historically, the deficiency distance is a generalization of the older concept of
 sufficiency ordering of experiments.

\begin{defin}
\label{defin:sufficient} \textit{(Sufficient Experiments
\cite{Blackwell_Ordering},\cite{Torgersen})} Experiment $\mathcal{E}
= (\mathcal{X}; P_{\theta}, \theta \in \Theta)$ is sufficient for
experiment $\mathcal{F} = (\mathcal{Y}; Q_{\theta}, \theta \in
\Theta)$ iff $\delta(\mathcal{E},\mathcal{F}) = 0$. The ordering of
experiments based on sufficiency is also denoted by $\mathcal{E}
\geq_{S} \mathcal{F}$.
\end{defin}

Note that ordering $\geq_{S}$ is a partial ordering on the set of
experiments with the same parameter set. This ordering is sometimes
also referred to as Blackwell ordering. We next discuss the basic
properties of the deficiency distance.

\begin{theorem}
\textit{(Properties of Deficiency Distance)} Let
$\mathcal{E},\mathcal{F},\mathcal{G}$ be experiments. Then

(i) $0 \leq \delta(\mathcal{E},\mathcal{F}) \leq 2 - 2 (\# \Theta)^{-1}$

(ii) $\delta(\mathcal{E},\mathcal{G}) \leq \delta(\mathcal{E},\mathcal{F}) +
\delta(\mathcal{F},\mathcal{G})$

iii) $\delta(\mathcal{E},\mathcal{E}) = 0$

\end{theorem}
\textit{Proof:} Theorem 6.2.24 in~\cite{Torgersen}. $\Box$

A special case of a statistical experiment is a linear experiment
defined below\footnote{A more general definition of linear
experiment may be given as in~\cite{Torgersen}, Section 8. }:

\begin{defin} \textit{(Linear Normal Experiment)}
Linear normal experiment $\mathcal{E}$ is denoted as
$\mathcal{E}(A,\Sigma; \beta \in \mathcal{R}^n)$ and is experiment
with the sample distribution $N(A'\beta, \Sigma)$ where the
parameter is $\beta \in \mathcal{R}^{n}$ and the parameter set is
$\mathcal{R}^{n}$. Real valued matrices $A$ and $\Sigma$ are assumed
to be known.
\end{defin}

Two linear experiments can be compared as follows.

\begin{theorem}
\textit{(Linear Experiment Comparison)}
\label{theorem:linear_comparison} (a) Linear normal experiment
$\mathcal{E}(A,\Sigma; \beta \in \mathcal{R}^n)$ is sufficient for
$\mathcal{F}(B,\Gamma; \beta \in \mathcal{R}^n)$ (i.e.
$\delta(\mathcal{E},\mathcal{F}) = 0$) if
\begin{equation}
A \Sigma A' - B \Gamma B'  \text{ is non-negative definite matrix.}
\end{equation}
(b) Let $\Sigma = I$ and $\Gamma = I$ then
$\delta(\mathcal{E},\mathcal{F}) < 2$ iff $\textit{range}(B)
\subseteq \textit{range}(A)$. Furthermore, if $\textit{range}(B)
\subseteq \textit{range}(A)$ then
\begin{equation}
\delta(\mathcal{E},\mathcal{F}) = \|N(0, (B'(A A')^{+} B) \vee I) -
N(0, I)\|.
\end{equation}
\end{theorem}
\textit{Proof:} Theorem 8.2.13 and 8.5.7 in~\cite{Torgersen}. $\Box$

General statistical results and inequalities can provide important
insights into the operation and efficient design of a communication
system. An early result on the influence of the communication
channels based on Blackwell ordering of channel transfer matrices
has been presented~\cite{Shannon_Ordering}. More recently,
monotonicity results on the influence of Rician fading on the
capacity of MIMO systems has been demonstrated
in~\cite{Lapidoth_Monotone}. Based on general inequalites of
stochastic majorization, performance comparison of various receivers
in multipath fading channels and the influence of power delay
profile has been shown in~\cite{Djonin_Power_Profile}.

Starting with the general notion of deficiency, we will present in
the next section the application of these concepts to the comparison
of error correction codes. Namely, in
Subsection~\ref{sec:linear_comparison} we present the block error
correction code comparison in additive white Gaussian Noise (AWGN)
channels based on linear experiments and block-error rates. In
Subsection~\ref{sec:finite_comparison} we present the
 bit-error rate comparison of error-correction codes used in discrete channels
 based on matrix majorization.

%
%
\section{Error Correction Block Code Comparison}\label{sec:code_comparison}
%
%

Let $(M,n)$ block error correction code denoted with $\mathcal{C}$
be defined as a map $\phi_{\mathcal{C}}: i \rightarrow x_i$ from the
set $i \in  \Theta = \{1,\ldots,M\}$ of possible information
messages to the set of codewords $x_i \in \mathcal{A}^{n}$  for $i =
1,\ldots,M$. Let $\mathcal{A}$ denote the alphabet of the code
symbols.

Let the received message be $z$ if the transmitted code word is $x = \phi_{\mathcal{C}}(y)$ as a result
of encoding the information message $y$. The dependence of received message $z$ on $x$ is described with a probabilistic law,
examples of which will be discussed in more detail in the following sections.

The original transmitted message $y$ is recovered through the usage of \textit{decoder}
$\delta$ which maps the received message $z$ to a possible transmitted data message
$y$
\begin{equation}
\delta(z) \rightarrow y.
\end{equation}
In the information theory as well as communications practice, a
common metric used to evaluate the performance of the block error
correction code is the packet (code-word) error probability. Let the
transmitted message be $y$. Then, the packet error probability of
code $\mathcal{C}$ is equal to
\begin{equation}
P_{e}^{\mathcal{C}}(y) = \Exp \left[ I_{\{\delta(z) \neq y\} }
\right]
\end{equation}
where expectation is over realizations of the random variable $z$ of the
received code word given transmitted data message $y$.
%
%
\subsection{Comparison of Error Correction Codes in AWGN Channels Based on
Linear Normal Experiments} \label{sec:linear_comparison}
%
%

In this section we will address the transmission of the coded
message over an additive white Gaussian noise channel. Namely,
assume that code word $x \in \mathcal{R}^{n}$ is being transmitted
and that the alphabet of the code symbols is $\mathcal{R}$. The code
words are assumed to be energy bounded such that $x x' \leq E$ where
parameter $E$ is the upper bound on the energy of the code word. The
received data is given as
\begin{equation}
z = x + \nu
\end{equation}
where $\nu$ is the additive Gaussian noise with covariance matrix $E
[\nu \nu'] = \sigma^{2} I$. The signal to noise ratio is therefore
upper bounded with $\frac{E}{n \sigma^{2}}$.

For transmission in  additive white Gaussian noise channel, a block
error correction code can be interpreted as a linear experiment
$\mathcal{E}(A,I \sigma^2;\beta \in \mathcal{R}^{M})$ such that the
parameter set is $\Theta = \mathcal{R}^{M}$ and matrix is $A = [x_1
x_2 \ldots x_{M} ]'$. The $i$-th information message is represented
with the parameter $\beta = e_i, i = 1,...M$. Matrix $A$ that
uniquely describes such a block code will be called the \textit{code
matrix} and the adjoint block code will be denoted in short with
$\mathcal{A}$. In the context of the comparison of experiments,
decoder $\delta$ can be interpreted as the decision rule, while
$I_{\{\delta(z) \neq \beta\} }$ is the loss function, for a received
message $z$ and transmitted data message $\beta$. A comparison of
error-correction code and statistical experiments terminology is shown in
Table~\ref{table:terminology_comparison} for easy reference.

\begin{table}
 \caption{Equivalent terminology used in statistical experiments and
error-correction codes}
 \label{table:terminology_comparison}
\begin{center}
\begin{tabular}{|c|c|}
        \hline
Statistical Experiments & Error-Correction Codes \\
        \hline
        \hline
parameter $\theta$   & information message  \\
        \hline
decision rule  & decoding algorithm \\
        \hline
sample space   & set of received code words \\
        \hline
loss function  & rate-distortion measure \\
        \hline
\end{tabular}
\end{center}
\end{table}

\begin{prop} (\textit{Comparison of two codes based on the packet error probability}) \label{prop:code_comparison}
Let $A$ and $B$ be two code matrices that define block error
correction codes $\mathcal{A}$ and $\mathcal{B}$ respectively. (a)
If noise variance $\sigma^2$ is a known parameter and if
\begin{equation}
A A' - B B'  \text{ is non-negative definite matrix}
\end{equation}
then there exists a decoder of the code $\mathcal{A}$ that will
always have smaller bit error rate than any decoder of the code
$\mathcal{B}$
\begin{equation} \label{eq:PER_comparison}
P_{e}^{\mathcal{A}}(\beta) \leq P_{e}^{\mathcal{B}}(\beta)
\end{equation}
for any transmitted data message $\beta = e_{i}, i \in 1,\ldots,M$.

(b) If the above condition is not satisfied and if
$\textit{range}(B) \subseteq \textit{range}(A)$ then for any data message $\beta$
\begin{equation}
P_{e}^{\mathcal{A}}(\beta) - P_{e}^{\mathcal{B}}(\beta) \leq
\delta(\mathcal{A},\mathcal{B}) = \|N(0, (B'(A A')^{+} B) \vee I) -
N(0, I)\|
\end{equation}
where $P_{e}^{\mathcal{A}}(\beta)$ and $P_{e}^{\mathcal{B}}(\beta)$ are packet error rates of
block codes $\mathcal{A}$ and $\mathcal{B}$ respectively.
\end{prop}

\textit{Proof:} The proof follows from the definition of the
deficiency distance and Theorem~\ref{theorem:linear_comparison}. Let
$\Theta_{0} = \{e_i|i = 1,2,\ldots,M \}$ and the decision space be
$\mathcal{Y} = \{1,2,\ldots,M\}$, while the loss function be defined
as
\begin{equation}
\mathcal{L}(\theta,y) = I_{\{y \neq max_i \theta_i\}}, \theta = [\theta_1,\ldots,\theta_M] \in
\Theta_{0}, y \in \mathcal{Y}. \label{eq:loss_PER}
\end{equation}
Expected value of the above loss function corresponds to the risk
function that is equal to the block error probability when averaged
over random noise realizations, i.e,
\begin{equation}
P_e(x) = \int P_{\theta}(x) \mathcal{L}(\theta,\sigma(x)) dx.
\end{equation}
According to the Definition~\ref{defin:sufficient}, the statement
(a) of this proposition immediately follows from
Theorem~\ref{theorem:linear_comparison}(a). Similarly, (b) directly
follows from Theorem~\ref{theorem:linear_comparison}(b) where the
Definition~\ref{defin:deficiency} is applied for the loss function
in (\ref{eq:loss_PER}) and by noting that $\max_{y}L(\theta,y) =
1$.$\Box$

\begin{remark} (\textit{Influence of the Information Message Distribution})
Note that the conclusions of the Proposition~\ref{prop:code_comparison} are
valid for \textit{any} data message $\beta = e_{i}, i = 1,\ldots,M$. Therefore,
it follows that the code comparison results of
Proposition~\ref{prop:code_comparison} are valid for any a priori distribution
of data messages.
\end{remark}

\begin{remark} (\textit{Loewner Order and Moment Generating Matrices})
Matrix partial order $A \geq B$ whenever $A A' - B B'$ is
non-negative definite is commonly referred to as Loewner order. In
the experiment design literature matrix $A A'$ is commonly referred
as the moment matrix of a linear experiment $\mathcal{E}(A,\Sigma;
\beta \in \mathcal{R}^n)$.
\end{remark}

\begin{remark} (\textit{Bayes risk})
The conclusions of the Proposition~\ref{prop:code_comparison} are
valid for the comparison of two codes with respect to \textit{any}
loss function and not just the packet-error probability. Namely, any
Bayesian risk function can be used and the conclusions of the
Proposition still hold.
\end{remark}

\begin{remark}
(\textit{Limitations of the code comparison based on linear
experiments}) The code comparison based on Loewner order is very
general and strong. There are two reasons for this statement: (i)
with respect to the block code, parameter set of the adjoint linear
experiment is extended from a finite set to the $n$-dimension space
$\mathcal{R}^n$, and (ii) the comparison is valid for any risk
function and may encompass risk function that may not be of interest
in the code design and performance analysis. Therefore, for some
applications the packet error rate bound based on the deficiency
bound may be loose.
\end{remark}

Next, we give an example of the code comparison based on the linear
experiment comparison The codes compared have the same parameter
set.

\begin{example}
(a) BCH(63,7) code is better than BCH(15,7) code for any loss
function. (b) However, code comparison between BCH(31,7) and
BCH(15,7) cannot be established based on
Proposition~\ref{prop:code_comparison}. These conclusions are easily
established by calculating their respective code matrices and
checking if Loewner ordering holds.
\end{example}

%
%
\subsection{Comparison of Error Correction Codes in Discrete Channels: Matrix
Majorization and Zonotopes} \label{sec:finite_comparison}
%
%

In this section we assume that the code symbol alphabet
$\mathcal{A}$ is finite with $l$ elements. Further, we assume that
channel outputs also belong to the finite alphabet of $w$ symbols,
i.e. that the channel is discrete. In light of the experiment
comparison terminology, therefore $w^n$ possible received noisy
code-words are elements of the sample space $\mathcal{X}$, while $M
= 2^k$ information messages (of length $k$ bits) are elements of the
parameter set $\Theta$.

A finite code symbol alphabet block error correction code
$\mathcal{E}$ can be represented with a row-stochastic $2^k \times
l^n$ code matrix $M_{\mathcal{E}}$. The code word
$\phi_{\epsilon}(i)$ corresponding to the information message $i$ is
the element of the set $\mathcal{A}^{n}$. Row $i$ of
$M_{\mathcal{E}}$ corresponds to the information message $i \in
1,\ldots,2^k$ and is equal to $e_{\phi_{\mathcal{E}}(i)}$, i.e. it
is equal to the unity vector with $1$ in $\phi_{\mathcal{E}}(i)$-th
place. As will be demonstrated below using
Theorem~\ref{theorem:Majorization_Preservation}, the ordering of
information and code messages is arbitrary and our proceeding
results do not depend on this ordering.

Channel $\mathcal{C}$ is modeled as follows. Since the number of
observations is considered to be finite, the channel is modeled with
a row stochastic $l^n \times w^n$ matrix $C = [p_{ij}] \in
\mathcal{M}_{l^{n}w^{n}}$. Probability $p_{ij}$ is the probability
of receiving $i$-th element of the sample space $\mathcal{X}$ if
code word $j \in \mathcal{A}^{n}$ is being sent. This general
channel description incorporates binary symmetric channels (BSC) (as
discussed in the Example~\ref{example:Zonotope}) as well as many
other channels such as bursty error channels.

Following the naming conventions introduced in Section I, coding
experiment will be considered as the observation of the parameter
(or message) $\theta \in \Theta$ in channel $\mathcal{C}$ after
coding the message with the block code $\mathcal{E}$. Therefore, the
probability measure $P_{\theta}$ of such experiment on the finite
sample space $\mathcal{X}$ is given with the rows of the
\textit{transfer matrix} $M_{\mathcal{E}} C$. Using the theory of
experiment comparison, performance comparison of two block codes
$\mathcal{E}$ and $\mathcal{F}$ in the same channel $\mathcal{C}$
can be based on their transfer matrices $M_{\mathcal{E}} C$ and
$M_{\mathcal{F}} C$.

Let us consider the channel $\mathcal{C}$. Then, code $\mathcal{E}$
is sufficient for code $\mathcal{F}$ if $M_{\mathcal{E}} C M =
M_{\mathcal{F}} C$, for some $w^n \times w^n$ dimensional
row-stochastic matrix $M$~\cite{Torgersen}. This condition is
equivalent to the matrix majorization \cite{dahl} of matrix
$M_{\mathcal{E}} C$ with respect to $M_{\mathcal{F}} C$ and is also
denoted with $M_{\mathcal{E}} C \succ M_{\mathcal{F}} C$. Therefore,
for any loss function and decoder associated with decoding of the
code $\mathcal{F}$, there exists a decoder for the code
$\mathcal{E}$ that produces less or equal risk. In addition, under
the above majorization condition, code $\mathcal{E}$ can have
smaller code word error probability than the code $\mathcal{F}$.

Several properties of the matrix majorization are shown
in\cite{dahl}. This article also demonstrates that it is possible to
check the matrix majorization ordering between two matrices by
checking the feasibility of a linear program.

%
%

It is stated that the linear operator $T: \mathcal{M}_{nm}
\rightarrow \mathcal{M}_{nm}$ preserves the matrix majorization
ordering if $A \succ B \implies T(A) \succ T(B)$ for some $A,B \in
\mathcal{M}_{nm}$.

\begin{theorem}{(\em Preservation of the Matrix Majorization Ordering \cite{Majorization_Preservation})}
\label{theorem:Majorization_Preservation} A linear operator $T:
M_{nm} \rightarrow M_{nm}$ preserves the matrix majorization
ordering if and only if $T(X) = L X P$, where $L \in M_{nn}$ is an
invertible matrix and $P \in  M_{mm}$ is a permutation matrix.
\end{theorem}

An interpretation of the
Theorem~\ref{theorem:Majorization_Preservation} in the context of
the error correction code comparison is as follows. Since, matrix
majorization is preserved if matrices are multiplied from the right
by any permutation matrix $P$, it is obvious that code comparison
ordering is preserved for any permutation of the received code words
in the transmission channel. Also, let us first consider the case
when invertible matrix $L$ in the above theorem is also a
permutation matrix. Then, it follows directly from
Theorem~\ref{theorem:Majorization_Preservation} that code comparison
ordering is preserved for any permutation of the input data
messages. The general case  of any invertible matrix $L$ is not of
interest as it would amount to randomization of data messages at the
input of the encoder.

%
%

Due to the large dimensions of the matrices $M_{\mathcal{E}}C$ and
$M_{\mathcal{F}}C$ it might not computationally be simple to check
if matrix majorization ordering $\succ$ can be established between
these two transfer matrices. Therefore, to simplify the setting and
to provide more insight into the matrix majorization ordering we
will be considering next the special case of detecting a single bit
of the information message $\theta$ consisting of $k$ bits.

To accomplish that, we first have to introduce our following assumption:

\begin{assum}
The a priori probability distribution of data messages $p(\theta)$ is known.
\end{assum}

This assumption is warranted in practical systems. For example,
source coding is usually used prior to error correction coding which
renders information messages uniformly distributed.

Let us concentrate on the decoding of the $r$-th bit of the $k$-bit
long information message $\theta = \left[ b_{1} b_{2} \cdots b_{r}
\cdots b_{k} \right]$. Therefore bit $b_{r}$ is the parameter of the
experiment and the parameter set is $\Theta = \{0,1\}$. All other
bits in the information message are to be considered to be nuisance
parameters. In the long term, the effect of the nuisance parameters
can be averaged out by introducing the \textit{equivalent code
matrix} $M_{\mathcal{E}}^{r}$ as follows. Let $\Theta^{r}(0) =
\{\theta = \left[ b_{1} b_{2} \cdots b_{r} \cdots b_{k} \right] |
b_{r} = 0 \}$ be the set of information messages for which the
$r$-th bit is equal to $0$. Similarly, let $\Theta^{r}(1) = \{\theta
= \left[ b_{1} b_{2} \cdots b_{r} \cdots b_{k} \right] | b_{r} = 1
\}$ be the set of information messages for which the $r$-th bit is
equal to $1$.

The equivalent $2 \times w^n$ transfer matrix
$\tilde{M}^{r}_{\mathcal{E}}$ for $r$-th bit will be defined as
\begin{equation}
\tilde{M}^{r}_{\mathcal{E}} = \left[
\begin{array}{c} \sum_{\theta \in \Theta^{r}(0)}
p(\theta) M_{\mathcal{E}}(\theta, : ) \\
\sum_{\theta \in \Theta^{r}(1)} p(\theta) M_{\mathcal{E}}(\theta, :
)
\end{array}
\right].
\end{equation}
The coefficients of this equivalent transfer matrix will be denoted
with $\tilde{e}_{jk}^{r}; j = 1,2; k = 1,\ldots,w^n$.

By introducing the equivalent experiment for each of the bits of the
information message $\theta$, we can represent the block code
$\mathcal{E}$ as a sequence of $k$ experiments with parameter set of
just two elements. In the statistical literature the experiment with
the parameter set of cardinality 2 is usually called a {\em
dichotomy}. Dichotomy will be denoted with $\mathcal{D} = \{D,
\theta \in \{0,1\} \}$, where $D$ is a 2-row stochastic transfer
matrix. In the case of dichotomies, the matrix majorization has
several very useful simplifying properties. However, to be able to
use these properties, we first have to introduce the concepts of
zonotopes.

\begin{defin}
{\em (Dichotomy Zonotopes~\cite{dahl})} Consider a dichotomy
$\mathcal{D}_{A} = \{A, \theta \in \{0,1\} \}$ where $A = \{a_{ij}
\}$ is a $2 \times n$ dimensional row stochastic matrix. The
zonotope of the dichotomy $\mathcal{D}_{A}$ is defined as
\begin{equation}
Z(\mathcal{D}_{A}) = \sum^{\oplus}_{i \in \{1,\ldots,n\} }
\textrm{conv}(0,A(:,i))
\end{equation}
where the addition $\oplus$ in the previous equation is the Minkowski addition.
\end{defin}
Recall that Minkowski addition $\oplus$ of two sets is the set of sums of all possible
combination of elements from these two sets, i.e.
\begin{equation}
\mathcal{A} \oplus \mathcal{B} = \{ a+b | a \in \mathcal{A}, b \in \mathcal{B} \}
\end{equation}
Dahl~\cite{dahl} showed that the dichotomy zonotope is a polygone
that contains the origin and is symmetric with respect to the point
$(\frac{1}{2},\frac{1}{2})$. The upper boundary of the dichotomy
zonotope can be calculated as
\begin{eqnarray}
\beta_{\mathcal{A}}(x) &=& \max_y \{y| (x,y) \in Z(\mathcal{A})\} \\
                       &=& \max \{ \sum_{j=1}^{n} a_{2j} \delta_j |
                           \sum_{j=1}^{n} a_{1j} \delta_j \leq x, 0 \leq \delta_j \leq 1\}
\end{eqnarray}
for $x \in [0,1]$. Now, two dichotomies can be compared as follows:
\begin{theorem} (\textit{Comparison of Dichotomies~\cite{dahl}})
Let $\mathcal{A} = \{A, \theta \in \{0,1\} \}$ and $\mathcal{B} =
\{B, \theta \in \{0,1\} \}$  be two dichotomies with the same
parameter set. Then, the following properties are equivalent:

(a) $\mathcal{A} \leq_{S} \mathcal{B}$

(b) $A \prec B$

(b) $Z(\mathcal{E}) \subseteq Z(\mathcal{F})$

(c) $\beta_{\mathcal{A}}(x) \leq \beta_{\mathcal{B}}(x), x \in
[0,1]$.
\end{theorem}

\textit{Proof:} Corollary 4.2 in~\cite{dahl}. $\Box$

Now, using the per-bit equivalent matrix representation of the block
code experiment we can state the following corollary regarding the
bit-error probabilities of the decoding of a particular bit of a
code word.
\begin{corol} (\textit{Per-bit Code Comparison})
Let $\mathcal{E}$ and $\mathcal{F}$ be two block error correction
codes used in channels $\mathcal{C}_{1}$ and $\mathcal{C}_{2}$ with
equivalent transfer matrices $\tilde{M}_{\mathcal{E}}^{r} =
\{\tilde{e}^r_{jk}\}$ and $\tilde{M}_{\mathcal{F}}^{r} =
\{\tilde{f}^r_{jk}\}$, respectively. Then, the probability of
decoding the $r$-th bit of the information message of code
$\mathcal{E}$ can always be less than that of decoding code
$\mathcal{F}$ if
\begin{eqnarray}
&\max & \{ \sum_{j=1}^{n} \tilde{e}^{r}_{2j} \delta_j |
\sum_{j=1}^{n}
\tilde{e}^{r}_{1j} \delta_j \leq x, 0 \leq \delta_j \leq 1\} \geq \\
&\max & \{ \sum_{j=1}^{n} \tilde{f}^{r}_{2j} \delta_j |
\sum_{j=1}^{n} \tilde{f}^{r}_{1j} \delta_j \leq x, 0 \leq \delta_j.
\leq 1\}
\end{eqnarray}
\end{corol}
We next illustrate the shape and certain properties of the zonotope of a block
error correction code.
\begin{example} (\textit{BCH Code Comparison using Zonotopes}) \label{example:Zonotope}
Consider the use of Hamming(15,7) code $\mathcal{E}$ in a binary
symmetric channel (BSC) with probability of error $p$. In Figure 1,
we show zonotopes for $Z_1(\mathcal{E})$ and $Z_2(\mathcal{E})$ for
the use of this code in BSC with probabilities of error $p_1 = 0.1$
and $p_2 = 0.2$, for the first information bit. It is obvious that
the $Z_1(\mathcal{E}) \supset Z_2(\mathcal{E})$ and that code
$\mathcal{E}$ will be better performing in the channel with error
probability $p_1$, than in the channel with error probability $p_2$.
\end{example}
\begin{figure}[!tbh]
\centering             
\includegraphics[angle=0,width=6.75in,height=4.75in]{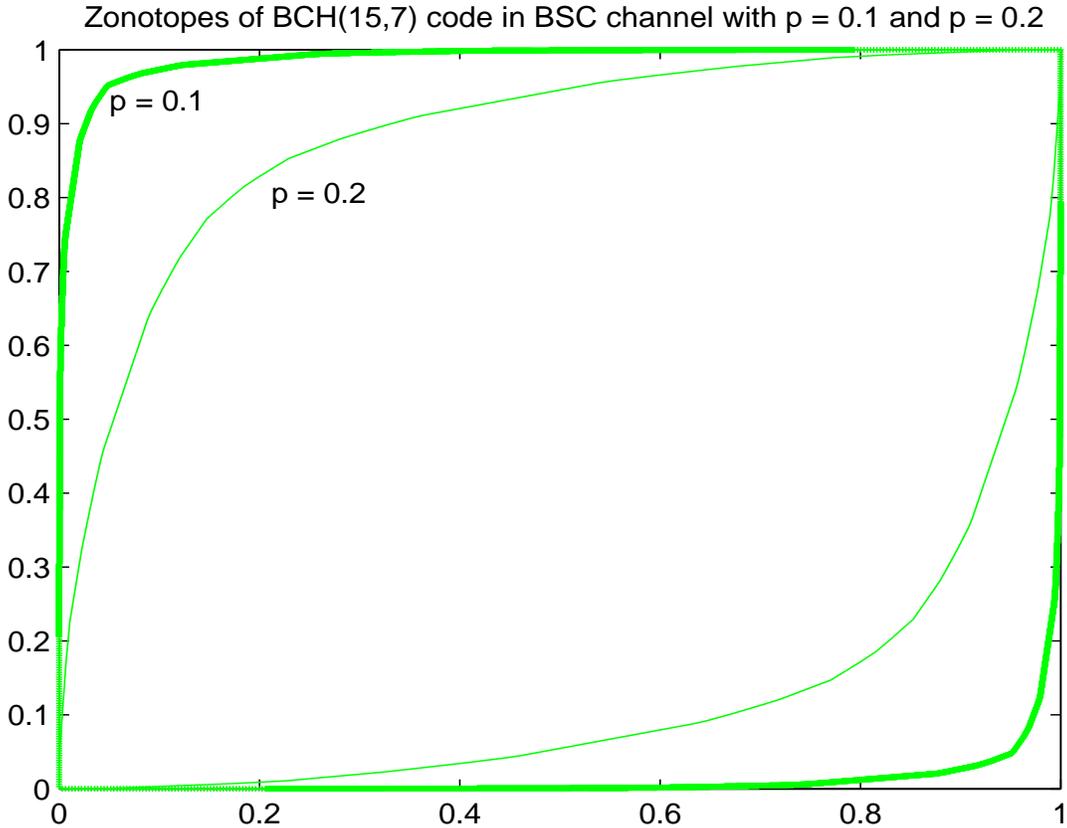}
\caption{Illustration of the zonotopes of the BCH(15,7) codes in the
BSC channel with $p = 0.2$ and $p = 0.1$.} \label{figure:Policy3D}
\end{figure}
\begin{remark}
The error-correction comparison based on dichotomies and bit-error
rates is more specific than the comparison based on the comparison
of the linear experiments. This is due to the fact that comparison
based on dichotomies is using more information about the structure
of the parameter set to be used to convey information. As opposed to
that the parameter set in the case of linear experiments is the set
$\mathcal{R}^{k}$ even if the true set of information messages is
finite set $\Theta = \{0,\ldots,M\}$.
\end{remark}

%
%
\section{Extensions and Conclusion}

%
%
In a similar manner one can compare the performance of the spreading
codes in CDMA systems or compare performances of two distinct
Multiple-Input Multiple-Output (MIMO) channels. For example,
consider a MIMO channel
\begin{equation}
y = A x + n \label{eq:CDMA}
\end{equation}
where $x$ is a $t$-dimensional column vector of the transmitted message, $y$ is
a $r$-dimensional column vector of the received message, $A$ is $t \times
r$-dimensional channel matrix, and $n$ is the additive Gaussian noise with
covariance matrix $I_r$.

It is obvious that results of the
Theorem~\ref{theorem:linear_comparison} can be directly applied and
the channels can be compared using the concepts of Loewner ordering
and deficiency. For example, MIMO channel defined with channel
matrix $A$ is better than MIMO channel defined with channel matrix
$B$ iff $A A' - B B'$ is non-negative definite. The criterion for
performance comparison can be any loss function for the estimation
of the uncoded transmitted signal $x$. Results with a similar flavor
have been obtained in~\cite{Lapidoth_Monotone} for the comparison of
two Rician MIMO channels but without the use of powerful theory of
the comparison of experiments.

In conclusion, let us mention that the elegant statistical results on the comparison of experiments and
deficiencies~\cite{Blackwell_Ordering},~\cite{Torgersen},~\cite{LeCam}
have been known for several decades. However, to the best knowledge
of the author, despite their appeal to the problems of information
transfer these results have received limited attention in the
information and communication theory before.

\bibliographystyle{IEEEtran}

\bibliography{WorkBibliography}

%
\newpage

\end{document}